# Matching Interdiction

Rico Zenklusen

Institute for Operations Research, ETH Zurich

rico.zenklusen@ifor.math.ethz.ch

May 8, 2019

**Abstract**

In the matching interdiction problem, we are given an undirected graph with weights and interdiction costs on the edges and seek to remove a subset of the edges constrained to some budget, such that the weight of a maximum weight matching in the remaining graph is minimized. In this work we introduce the matching interdiction problem and show that it is strongly NP-complete even when the input is restricted to simple, bipartite graphs with unit edge weights and unit interdiction costs. Furthermore, we present a pseudo-polynomial algorithm for solving the matching interdiction problem on graphs with bounded treewidth. The proposed algorithm extends the approach that is typically used for the creation of efficient algorithms on graphs with bounded treewidth to interdiction problems.

## 1 Introduction

We are interested in the problem of minimizing the maximum weight of matchings in a given graph by removing edges constrained to some interdiction budget. This problem is inspired by other interdiction problems as the network flow interdiction problem where one tries to minimize the maximum flow of a given network by removing arcs constrained to a budget (c.f. [9, 11, 12]). One motivation for studying this type of problems is to get a robustness measure for solutions to combinatorial optimisation problems based on maximum matchings. To the best of the authors knowledge, the matching interdiction problem has not been studied so far.

In [6], a problem related to matching interdiction was studied with a focus on graph-theoretical aspects, namely the one of finding a minimum $d$-blocker in a given graph. The task is to determine a subset of the edges of minimum cardinality such that their removal from the graph decreases the cardinality of a maximum matching by at least $d$ units. In [5] a polynomial delay algorithm for finding all minimum 1-blockers of a bipartite graph, that contains a perfect matching, is presented.

In a first part of this work, we present some hardness results for the matching interdiction problem. In particular we give a reduction from the knapsack problem showing that matching interdiction is $NP$-complete even when restricted on graphs consisting only of isolated edges, i.e., every connected components of the graph consists of a single edge. Additionally, we show that the problem is strongly $NP$-complete for simple, bipartite graphs with unit interdiction costs and unit edge weights.

We furthermore present a pseudo-polynomial algorithm for the matching interdiction problem on graphs with bounded treewidth using dynamic programming. Numerous combinatorial problems that are hard in general can be solved efficiently on graph with bounded treewidth by a rather standard approach [1, 4, 8]. However, these algorithms deal with problems that do not have a min-max character as it is the case with interdiction problems. The proposed algorithm extends the approach that is typically used for the creation of efficient algorithms on graphs with bounded treewidth to interdiction problems. The presented approach can be adapted to solve other interdiction problems on graphs with bounded treewidth.

The paper is organized as follows. We begin by giving some definitions and notations in Section 2. In Section 3, hardness results for the matching interdiction problem are presented. Section 4 presents a pseudo-polynomial algorithm for solving the matching interdiction problem on graphs with bounded treewidth.

## 2 Definitions and Notations

Let $G = (V, E)$ be an undirected graph. Edges are represented as unordered pairs of vertices. For $V' \subseteq V$, we denote by $G[V']$ the subgraph of $G$ induced by $V'$, i.e., $G[V'] = (V', E \cap (V' \times V'))$. Furthermore for $E' \subseteq E$ we denote by $G - E'$ the subgraph $(V, E \setminus E')$. A graph is called simple, if it contains neither loops nor parallel edges. For $v \in V$, we denote by $d(v)$ the degree of $v$, i.e., $d(v) = |\{e \in E \mid v \in e\}|$. A bipartite graph $G$ is denoted by $G = (X, Y, E)$ where $E$ is the set of edges and $X, Y$ is a bipartition of the vertices of $G$ such that $E \subset X \times Y$. For a positive integer $k$ we denote by $C_k$ an undirected cycle over $k$ vertices. For some set $X$, we denote by $\mathcal{P}(X)$, the set of all of its subsets. For graph-theoretical terms used in this paper and not further specified in this section we refer to [10].

### 2.1 Matching interdiction

A *matching interdiction network* is a triple $(G, w, c)$, where $G = (V, E)$ is an undirected graph with edge weigths $w : E \to \mathbb{N}$ and edge interdiction costs $c : E \to \mathbb{N}$. Let $(G, w, c)$ be a matching interdiction network. We denote by $\mathcal{M}(G) \subseteq \mathcal{P}(E)$ the set of all matchings in $G$ and by $\nu(G)$ the weight of a maximum weight matching in $G$, i.e., $\nu(G) = \max_{M \in \mathcal{M}(G)}\{w(M)\}$. When necessary, we also use the notation $\nu(G, w)$ to specify the used weights. A set $U \subseteq E$ is called an interdiction set with respect to the budget $B$ if $c(U) = \sum_{e \in U} c(e) \leq B$. The *matching interdiction problem* asks to find for some fixed budget $B$, an interdiction set $U$, that minimizes $\nu(G - U)$. We use the notation $\nu_B(G) = \min\{\nu(G - U) | U \subseteq E, c(U) \leq B\}$ or equivalently $\nu_B(G, w, c)$ to specify the weights and costs defined on the edges $E$. For an undirected graph $G = (V, E)$ and $B \in \{0, \ldots, |V|\}$ we denote by $\nu_B^u(G)$ the value $\nu_B(G, w^u, c^u)$, where $w^u$ and $c^u$ are the unit weight respectively unit interdiction cost function, i.e., $w^u(e) = c^u(e) = 1 \ \forall e \in E$.

### 2.2 Treewdith and tree decompositions

A *tree decomposition* of a graph $G = (V, E)$ is a pair $(\mathcal{X} = \{X_i \mid i \in I\}, T = (I, F))$ with $T = (I, F)$ a tree and each node $i \in I$ has associated to it a subset of vertices $X_i \subseteq V$, such that

- $\cup_{i \in I} X_i = V$.

- For all edges $\{v, w\} \in E$ there exists an $i \in I$ with $\{v, w\} \subseteq X_i$.

- For all vertices $v \in V$, the set of nodes $\{i \in I \mid v \in X_i\}$ induces a subtree of $T$.

The *width* of the tree decomposition $(\mathcal{X}, T)$ is $\max_{i \in I}\{|X_i| - 1\}$. The *treewidth* of a graph $G$ is the minimum width over all tree decomposition of $G$. A graph with treewidth at most $k$ is also called a *partial $k$-tree*. This notion comes from an alternative definition of graphs with bounded treewidth [3]. For a given graph $G = (V, E)$ with treewidth bounded by a constant $k$, a tree decomposition with width $k$ and $|I| = O(|V|)$ can be found in linear time [2]. A tree decomposition $(\mathcal{X}, T)$ of $G = (v, E)$ is called *nice* if the tree $T$ is rooted and binary, and the nodes are of four types:

- *Leaf nodes* $i \in I$ are leaves of $T$ and satisfy $|X_i| = 1$.

- *Introduce nodes* $i \in I$ have one child $j$ with $X_i = X_j \cup \{v\}$ for some vertex $v \in V$.

- *Forget nodes* $i \in I$ have one child $j$ with $X_i = X_j \setminus \{v\}$ for some vertex $v \in V$.

- *Join nodes* $i \in I$ have two children $j_1, j_2$ with $X_i = X_{j_1} = X_{j_2}$.

A tree decomposition can easily be converted (in linear time) into a nice tree decomposition of the same width and with a linear growth in size [8]. Design and analysis of algorithms is often easier when dealing with nice tree decompositions.

## 3 Complexity

In this section, we present various hardness results for the matching interdiction problem on different input classes. The following natural decision version of the matching interdiction problem will be used for complexity analysis.

**MINT$(G, w, c, B, K)$**
Input: Matching interdiction network $(G, w, c)$, $B, K \in \mathbb{N}$.
Question: Decide whether $\nu_B(G) \leq K$.

The following theorem shows that even on very restricted graph classes, the MINT problem is NP-complete.

**Theorem 1.** *The MINT problem is NP-complete on the class of graphs consisting only of isolated edges.*

*Proof.* The MINT problem clearly lies in $NP$. To prove the hardness we use a reduction from the knapsack problem which is well known to be NP-complete (c.f. [7]). Consider an instance of a knapsack problem where we are given a finite set $I$, two positive integers $K, Z$ and for each $i \in I$ a size $s(i) \in \mathbb{N}$ and a value $v(i) \in \mathbb{N}$. The task is to decide whether there exists some set $I' \subset I$ with $\sum_{i \in I'} s(i) \leq Z$ and $\sum_{i \in I'} v(i) \geq K$.

Let $G = (V, E)$ be an undirected graph consisting of $|I|$ isolated edges $E = \{e_i \mid i \in I\}$. We define edge weights $w$ on $E$ that are equal to the values of the corresponding knapsack elements, i.e., $w(e_i) = v(i)$ for $i \in I$, and costs $c$ on $E$ that are equal to the sizes of the corresponding knapsack elements, i.e., $c(e_i) = s(i)$ for $i \in I$. Since every subset of $E$ is a matching in $G$, we have that the problem MINT$(G, w, c, Z, \nu(G) - K)$ reduces to finding a set $I' \subseteq I$, with $\sum_{i \in I'} c(e_i) = \sum_{i \in I'} s(i) \leq Z$ and $\sum_{i \in I \setminus I'} w(e_i) = \nu(G) - \sum_{i \in I'} v(i) \leq \nu(G) - K$. This is exactly the knapsack problem. $\square$

To show strong NP-completeness of the MINT problem, we will show that the problem remains NP-complete when restricted to unit edge weights and unit edge interdiction costs, i.e., NP-completeness of the following problem will be proven.

**MINTU($G, B, K$)**
Input:      Undirected graph $G = (V, E)$ and $B, K \in \{0, \ldots, |E|\}$.
Question:   Decide whether $\nu_B^u(G) \leq K$.

Even though the cost of every edge in a MINTU problem is set to one, higher costs can be modelled by adding parallel edges. In particular by adding to an edge $e \in E$, $|E|$ additional parallel edges, we can model an edge as *non-removable* since $B \leq |E|$, i.e., we have shown that the following problem reduces to the MINTU problem.

**NMINTU($G, N, B, K$)**
Input:      Undirected graph $G = (V, E)$, $N \subseteq E$ and $B, K \in \{0, \ldots, |E|\}$.
Question:   Decide whether there is a set $U \subseteq E \setminus N$, $|U| \leq B$ and $\nu(G - U) \leq K$.

The hardness of the MINTU problem will be shown by proving that NMINTU is NP-hard. To obtain complexity results for the MINTU problem that are valid for simple graphs we consider an alternative reduction from the NMINTU problem to the MINTU problem.

**Theorem 2.** *The problem* NMINTU($G, N, B, K$), *with $G$ a simple graph, can be polynomially reduced to* MINTU($G', B, K'$), *with $G'$ a simple graph.*

*Proof.* Let $G = (V, E)$ be an undirected graph, $N \subset E$ be a set of non-removable edges and $K, B \in \{0, \ldots, |E|\}$ as used as input in the NMINTU problem. We replace each edge $\{u, v\} \in N$ by the following construction, which we call a $|E|$-gadget (between $u$ and $v$): we add a complete bipartite graph $(X, Y, \widetilde{E})$ with $|X| = |Y| = |E| + 1$ and edges are added to link $u$ to all vertices in $X$ and $v$ to all vertices in $Y$ (see Figure 1). The vertices $u$ and $v$ are called the endpoints of the $|E|$-gadget. Let $G' = (V', E')$ be the graph obtained by this construction. The problem NMINTU($G, N, B, K$) is equivalent to MINTU($G', B, K + |N|(|E| + 1)$) because of the following observations.

Let $H$ be one of the $|E|$-gadgets in $G'$. Notice that we have $\nu(H) = |E| + 2$ and that there are $|E| + 1$ disjoint perfect matchings in $H$, all saturating both endpoints of $H$. Furthermore there are $|E| + 1$ disjoint matchings in $H$ with cardinality $|E| + 1$, that saturate none of the endpoints. We call this type of matchings *non-saturating maximum matchings*. When we have a matching $M'$ in $G'$ that does not use any edges of $H$ and we want to extend it with as many edges in $H$ as possible, there are two possibilities. Either at least one endpoint of $H$ is already saturated by $M'$. In this case the best we can do is to add a non-saturating matching of $H$ to $M'$, which increases the cardinality of the matching by $|E| + 1$. Or none of the endpoints of $H$ is saturated by $M'$. In this case we can add a perfect matching of $H$ to $M'$, thus increasing the cardinality of the matching by $|E| + 2$. Therefore, replacing an edge by a $|E|$-gadget simply increases the size of a maximum matching by $|E| + 1$. To get the equivalence between NMINTU($G, N, B, K$) and MINTU($G', B, K + |N|(|E| + 1)$), we finally show that removing any $|E|$ edges in $G$ from $|E|$-gadgets does not decrease the cardinality of a maximum matching in $G'$. This property is obtained by observing that since in every $|E|$-gadget there are $|E| + 1$ disjoint perfect matchings as well as $|E| + 1$ disjoint non-saturating maximum matchings, we have that after removing up to $|E|$ edges in $G'$ there is at least one matching of both types left for any $|E|$-gadget. We therefore have the desired property that the $|E|$-gadgets are "immune" to edge removals. □

The following lemma will later be used for proving NP-hardness of the NMINTU problem. A proof of the lemma can be found in the appendix.

**Lemma 1.** *Let $k \geq 4$ be an integer and let $G = (X, Y, E)$ be a simple bipartite graph such that*

(i) $|X| > k$;

(ii) $|Y| = \binom{k}{2}$;

(iii) $d(y) = 2$, $\forall y \in Y$ and $d(x) \geq 1$, $\forall x \in X$;

(iv) $G$ contains no $C_4$.

*Then $\nu(G) \geq k + 1$.*

**Theorem 3.** NMINTU *is NP-complete on simple bipartite graphs.*

*Proof.* The problem is clearly in NP. The theorem will be proven by a reduction from the CLIQUE problem which asks to find for a given simple undirected graph $H = (I, F)$ and some integer $r \leq |I|$, a clique in $H$ with cardinality $r$. The CLIQUE problem is well known to be NP-complete [7]. We construct a bipartite graph $G = (V, E)$ as follows: with each vertex $i \in I$, we associate a vertex $v_i \in V$ and with each edge $\{i, j\} \in F$ we associate a vertex $v_{ij} \in V$ (for notational convenience we set $v_{ij} = v_{ji}$); for each vertex $v_{ij} \in V$ we add a new vertex $\bar{v}_{ij}$ as well as an edge $\{v_{ij}, \bar{v}_{ij}\}$; finally for each edge $\{i, j\} \in F$, we add two edges $\{v_i, v_{ij}\}$ and $\{v_j, v_{ij}\}$. Furthermore, let $N = \{\{v_i, v_{ij}\} \mid \{i, j\} \in F\}$. See Figure 2 for an example of how $G$ is constructed from a given graph $H$.

Notice that the cardinality of a maximum matching $M$ in $G$ is $|M| = |F|$. Such a matching may be obtained by taking all the edges $\{v_{ij}, \bar{v}_{ij}\}$. We will now prove the following statement which finishes the proof: $H$ contains a clique of size $r$ if and only if the problem NMINTU$(G, N, \binom{r}{2}, \nu(G) - \frac{r(r-3)}{2})$ evaluates to true. Let us suppose that $H$ contains a clique $C$ of size $r$ and let $F_C \subseteq F$ be the edges of this clique. By defining $U = \{\{v_{ij}, \bar{v}_{ij}\} | \{i, j\} \in F_C\}$, we have $\nu(G - U) + \frac{r(r-3)}{2} = \nu(G)$. In fact a maximum matching in the graph $G - U$ is obtained by taking the remaining edges $\{v_{ij}, \bar{v}_{ij}\}$ (there are exactly $|F| - \frac{r(r-1)}{2}$ such edges) and the edges of a maximum matching in the subgraph induced by vertices $v_{ij}$ such that $\{i, j\} \in F_C$ and the vertices $v_i$ such that $i \in C$ (the cardinality of such a matching is at most $r$). Thus $\nu(G - U) \leq |F| - \frac{r(r-1)}{2} + r = |F| - \frac{r(r-3)}{2}$.

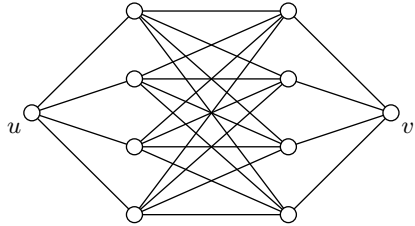

Figure 1: A 3-gadget between $u$ and $v$.

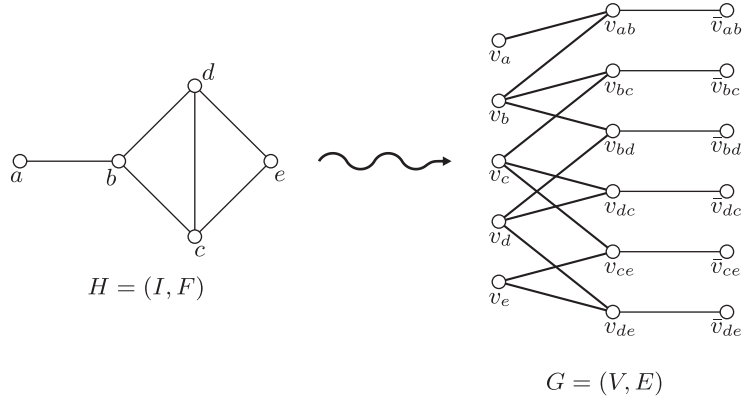

Figure 2: Example for the construction of the graph $G = (V, E)$ from a given graph $H = (I, F)$ with $I = \{a, b, c, d, e\}$. The edges in bold represent the set $N$.

Suppose now that there is a set $U \subseteq E \setminus N$ with $|U| = \binom{r}{2}$ and $\nu(G - U) \leq \nu(G) - \frac{r(r-3)}{2}$. By contradiction we furthermore assume that $H$ contains no clique of size $r$. Let $Y = \{v_{ij} \in V \mid \{v_{ij}, \bar{v}_{ij}\} \in U\}$ and $X$ be the subset of the vertices in $\{v_i \mid i \in I\}$ that are neighbors of $Y$ in $G$. The graph $G[X \cup Y]$ is a simple bipartite graph $\tilde{G} = (X, Y, \tilde{E})$ with the following properties:

(i) $|X| > r$, because $H$ is simple, does not contain a clique of size $r$, and has at least $\binom{r}{2}$ edges (the ones corresponding to the elements of $U$);

(ii) $|Y| = \binom{r}{2}$;

(iii) $d(v_{ij}) = 2, \forall v_{ij} \in Y$ and $d(v_i) \geq 1, \forall v_i \in X$;

(iv) $\tilde{G}$ contains no $C_4$ (since there are no multiple edges in $H$);

(v) $\nu(\tilde{G}) \leq r$.

Clearly these properties contradict Lemma 1. Thus there must be a clique of size $r$ in $H$ defined by the vertices $v_i, v_j$ such that $\{v_{ij}, \bar{v}_{ij}\} \in U$ and hence $|X| = r$. □

By combining Theorem 2 and Theorem 3 and observing that the reduction used in the proof of Theorem 2 preserves bipartition for a bipartite input graph, we get the following result.

**Theorem 4.** MINTU$(G, B, K)$ *is an* NP*-complete problem, even when the input graph $G$ is restricted to be bipartite and simple.*

## 4  Matching Interdiction on Graphs with Bounded Treewidth

In this section we present a pseudo-polynomial algorithm for solving the MINT problem on graphs with bounded treewidth. Let $k \in \mathbb{N}$ be a constant, $G = (V, E)$ be a graph with treewidth $k$ and $(\mathcal{X} = \{X_i \mid i \in I\}, T = (I, F))$ be a nice tree decomposition of $G$ with root $r \in I$, width $k$ and $|I| = O(|V|)$. As explained in Section 2 such a tree decomposition can be obtained in linear time. Furthermore, let $c : E \to \mathbb{N}$ and $w : E \to \mathbb{N}$ be an interdiction cost function and a weight function defined on the edges of $G$ and $B \in \mathbb{N}$ be the interdiction budget. Let $C = \sum_{e \in E} c(e)$. We associate with each node $i \in I$ a graph $G_i = (V_i, E_i)$, where $V_i$ is the union of all sets $X_j$ with $j$ equals $i$ or a descendant of $i$ in $T$ and $E_i = E \cap V_i \times V_i$ is the set of all edges in $E$ that have both endpoints in $V_i$.

Let $i \in I$, $b \in \{0, \ldots, B\}$ and $U_i \subseteq E_i$ with $c(U_i) \le b$ an interdiction set with budget $b$ in $G_i$. We are interested in the matchings in $G_i - U_i$ and how they can be completed to the graph $G$. The influence of $U_i$ on the rest of the graph is completely described when we know for each matching $M_i$ in $G_i - U_i$, its value and the vertices that are not saturated in $X_i$ by $M_i$. Keeping track of the non-saturated vertices in $X_i$, allows to complete the matching $M_i$ on the graph $G$ since $X_i$ are the vertices over which $G_i$ is connected with the rest of the graph $G$. Since we are interested in maximum matchings, it suffices to store for a given interdiction set $U_i$ a function $a_i^{U_i} : \mathcal{P}(X_i) \to \{0, \ldots, C\}$ such that $a_i^{U_i}(X)$ is the weight of a maximum matching in $G_i - U_i$ not saturating any vertex in $X$. Let $\mathcal{A}_i$ be the set of all functions from $\mathcal{P}(X_i)$ to $\{0, \ldots, C\}$. Note that $|\mathcal{A}_i| = (C+1)^{(2^{|X_i|})} \le (C+1)^{(2^{k+1})}$. We say that a function $a_i \in \mathcal{A}_i$ is *realized* by $U_i \subseteq E_i$ if $a_i = a_i^{U_i}$. Furthermore, an element $a_i \in \mathcal{A}_i$ is called *realizable* in $G_i$ with budget $b$ if there exists a set $U_i \subseteq E_i$ with $c(U_i) \le b$ that realizes $a_i$. Even though there are exponentially many interdiction sets in $G_i$, all we have to know from the subgraph $G_i$ is which functions in $\mathcal{A}_i$ are realizable. We denote by $\mathcal{A}_i^b$ the set of all functions in $\mathcal{A}_i$ that are realizable in $G_i$ with budget $b$. During the algorithm we will successively determine for all $i \in I$ and $b \in \{0, \ldots, B\}$, the set $\mathcal{A}_i^b$. It would suffice to keep track of only those elements $a_i \in \mathcal{A}_i^b$, that are efficient in the following sense. We say that $a_i \in \mathcal{A}_i^b$ is *efficient* if there is no element $a_i' \in \mathcal{A}_i^{b'}$ with $b' \le b$, $a_i'(X) \le a_i(X)$ $\forall X \subset X_i$ and $a_i' \ne a_i$. However, to simplify the presentation of the algorithm we keep track of all elements in $\mathcal{A}_i^b$. We represent the set $\mathcal{A}_i^b$ by a table $Q_i^b \in \{0,1\}^{\mathcal{A}_i}$, where $a_i \in \mathcal{A}_i^b$ will be encoded by $Q_i^b(a_i) = 1$. Notice that the size of the table $Q_i^b$ is equal to $|\mathcal{A}_i| = (C+1)^{(2^{|X_i|})}$.

We have that $\nu_B(G) \le K$ if and only if there is an entry $a_r \in \mathcal{A}_r^B$ with $a_r(\emptyset) \le K$. Therefore, when $Q_r^B$ is determined, we can answer the problem $\text{MINT}(G, w, c, B, K)$ by going through all elements $a_r \in \mathcal{A}_r$ that satisfy $a_r(\emptyset) \le K$ and test whether there is one such element with $Q_r^B(a_r) = 1$. If this is the case then $\nu_B(G) \le K$.

The proposed algorithm will compute the tables $Q_i^b$ in a bottom up order with respect to the nodes $i$, i.e., the tables $Q_i^b$ for $b \in \{0, \ldots, B\}$ will be computed after all tables of all descendants of $i$ are computed. In particular, the tables of the children of $i$ will be used to compute the tables corresponding to $i$. We will now discuss how the tables corresponding to a node $i$ are calculated. We distinguish four cases depending on whether the node $i$ is a leaf node, an introduce node, a forget node or a join node.

## 4.1 Leaf nodes

**Theorem 5.** *Let $i \in I$ be a leaf node. For any $b \in \{0, \ldots, B\}$ and $a_i \in \mathcal{A}_i$ we have*

$$Q_i^b(a_i) = \begin{cases} 1 & \text{if } a_i = \{0\}^{\mathcal{P}(X_i)} \\ 0 & \text{if } a_i \ne \{0\}^{\mathcal{P}(X_i)} \end{cases}.$$

*Proof.* Since $i$ is a leaf node we have $E_i = \emptyset$. Therefore for any $b \in \{0, \ldots, B\}$, the only possible interdiction set on the graph $G_i$ is the empty set and it realizes the element $\{0\}^{\mathcal{P}(X_i)}$. □

### 4.2 Introduce nodes

**Lemma 2.** *Let $i \in I$ be an introduce node with child $j$, $X_i = X_j \cup \{v\}$ and let $E_i^v = E_i \setminus E_j$, i.e., $E_i^v$ is the set of edges in $E_i$ adjacent to $v$. Let $U_i \subseteq E_i$ and we define $U_j = U_i \cap E_j$, $U_i^v = U_i \setminus E_j$ and denote by $a_i \in \mathcal{A}_i$ the element realized by $U_i$ in $G_i$ and by $a_j \in \mathcal{A}_j$ the element realized by $U_j$ in $G_j$. We have for $X \subseteq X_i$*

$$a_i(X) = \begin{cases} a_j(X) & \text{if } v \in X \\ \max\{a_j(X), \max_{\substack{\{v,w\} \in E_i^v \setminus U_i^v \\ \text{with } w \notin X}} \{c(\{v,w\}) + a_j(X \cup \{w\})\}\} & \text{if } v \notin X \end{cases}.$$

*Proof.* Let $\mathcal{M}_1 = \mathcal{M}(G_i[V_i \setminus X_i] - U_i)$. $\mathcal{M}_1$ can be partitioned into the matchings $\mathcal{M}_2 \subseteq \mathcal{M}_1$ that saturate $v$ and the matchings $\mathcal{M}_3 \subseteq \mathcal{M}_1$ that do not saturate $v$, i.e.,

$$\mathcal{M}_2 = \mathcal{M}(G_j[V_j \setminus X] - U_i),$$
$$\mathcal{M}_3 = \{\{v,w\} \cup M \mid \{v,w\} \in E_i^v, M \in \mathcal{M}(G[V_j \setminus (X_i \cup \{w\})] - U_i)\}.$$

Let $X \subseteq X_i$. We begin with the case $v \in X$. Since in this case there are no matchings in $\mathcal{M}_1$ that saturate $v$, i.e. $\mathcal{M}_3 = \emptyset$, we have $\mathcal{M}_1 = \mathcal{M}_2$ and hence

$$a_i(X) = \max\{c(M) \mid M \in \mathcal{M}_1\} = \max\{c(M) \mid M \in \mathcal{M}_2\} = a_j(X).$$

Otherwise if $v \notin X$ we have

$$\begin{aligned} a_i(X) &= \max\{c(M) \mid M \in \mathcal{M}_2 \cup \mathcal{M}_3\} \\ &= \max\{\max\{c(M) \mid M \in \mathcal{M}_2\}, \max\{c(M) \mid M \in \mathcal{M}_3\}\} \\ &= \max\{a_j(X), \max_{\substack{\{v,w\} \in E_i^v \setminus U_i^v \\ \text{with } w \notin X}} \{c(\{v,w\}) + \\ &\qquad \max\{c(M) \mid M \in \mathcal{M}(G_j[V_j \setminus (X \cup \{w\})] - U_j)\}\}\} \\ &= \max\{a_j(X), \max_{\substack{\{v,w\} \in E_i^v \setminus U_i^v \\ \text{with } w \notin X}} \{c(\{v,w\}) + a_j(X \cup \{w\})\}\}, \end{aligned}$$

proving the claim. □

As a consequence of Lemma 2 we obtain the following theorem which shows how the tables corresponding to an introduce node can be constructed when the tables of its child node are given.

**Theorem 6.** *Let $i \in I$ be an introduce node with child $j$ and let $X_i = X_j \cup \{v\}$. Let $E_i^v = E_i \setminus E_j$, i.e., $E_i^v$ contains all edges adjacent to $v$. For $b \in \{0, \ldots, B\}$ the table $Q_i^b$* *is correctly determined by the following algorithm.*

**foreach** $a_i \in \mathcal{A}_i$ **do**
$\quad | \quad Q_i^b(a_i) = 0;$
**end**
**foreach** $U_i^v \subseteq E_i^v$ with $c(U_i^v) \leq b$ **do**
$\quad$ **foreach** $a_j \in \mathcal{A}_j$ with $Q_j^{b-c(U_i^v)}(a_j) = 1$ **do**
$\quad\quad$ *An element $a_i \in \mathcal{A}_i$ is constructed as follows. For $X \subseteq X_i$ we set*

$$a_i(X) = \begin{cases} a_j(X) & \text{if } v \in X \\ \max\{a_j(X), \max_{\substack{\{v,w\} \in E_i^v \setminus U_i^v \\ \text{with } w \notin X}} \{c(\{v,w\}) + a_j(X \cup \{w\})\}\} & \text{if } v \notin X \end{cases}.$$

$\quad\quad Q_i^b(a_i) = 1;$
$\quad$ **end**
**end**

### 4.3 Forget nodes

**Theorem 7.** *Let $i \in I$ be a forget node with child $j$ and let $X_i = X_j \setminus \{v\}$. For $b \in \{0, \ldots, B\}$ the table $Q_i^b$ is correctly determined by the following algorithm.*

**foreach** $a_i \in \mathcal{A}_i$ **do**
$\quad | \quad Q_i^b(a_i) = 0;$
**end**
**foreach** $a_j \in \mathcal{A}_j$ with $Q_j^b(a_j) = 1$ **do**
$\quad$ *An element $a_i \in \mathcal{A}_i$ is constructed as follows. For $X \subseteq X_i$ we set*

$$a_i(X) = a_j(X) \ .$$

$\quad Q_i^b(a_i) = 1;$
**end**

*Proof.* Since $G_i = G_j$, we have in particular for any $X \subseteq X_i$ and $U_i \subseteq E_i$, $\mathcal{M}(G_i[V_i \setminus X] - U_i) = \mathcal{M}(G_j[V_j \setminus X] - U_i)$. Therefore, for each $U_i \subset E_i$ we have that the element $a_i \in \mathcal{A}_i$, that is realized by $U_i$ in $G_i$, and the element $a_j \in \mathcal{A}_j$, that is realized by $U_i$ in $G_j$, satisfy

$$a_i(X) = a_j(X) \quad \forall \, X \subset X_i,$$

implying the claim. $\square$

### 4.4 Join Nodes

To simplify notations, we define for $i \in I$ and $X \subseteq X_i$

$$\mathcal{S}_i(X) = \{(Y_1, Y_2) \mid Y_1, Y_2 \subseteq X_i, X \subseteq Y_1 \cap Y_2, Y_1 \cup Y_2 = X_i\} \ .$$

**Lemma 3.** *Let $i \in I$ be a join node with children $j_1, j_2 \in I$. Let $U_i \subseteq E_i$ and $U_{j_1}, U_{j_2}$ be the partition of $U_i$ defined by $U_{j_1} = U_i \cap V_{j_1}$ and $U_{j_2} = (U_i \cap V_{j_2}) \setminus U_{j_1}$. For $X \subseteq X_i$*

*we have*

$$\mathcal{M}(G_i[V_i \setminus X] - U_i) = \bigcup_{(Y_{j_1}, Y_{j_2}) \in \mathcal{S}_i(X)} \{M_{j_1} \cup M_{j_2} \mid M_{j_1} \in \mathcal{M}(G_{j_1}[V_{j_1} \setminus Y_{j_1}] - U_{j_1}),$$
$$M_{j_2} \in \mathcal{M}(G_{j_2}[V_{j_2} \setminus Y_{j_2}] - U_{j_2})\} \ .$$

*Proof.* The inclusion "⊇" follows by observing that since $T$ is a tree decomposition we have $E_{j_1} \cap E_{j_2} \subseteq X_i \times X_i$. We therefore have that $M_{j_1} \cup M_{j_2}$, as used in the right hand side of the equality, is effectively a matching in $G_i[V_i \setminus X] - U_i$.

To prove the inclusion "⊆" let $M \in \mathcal{M}(G_i[V_i \setminus X] - U_i)$. We define $M_{j_1} = M \cap E_{j_1}$, $M_{j_2} = (M \cap E_{j_2}) \setminus X_i \times X_i$, $Y_{j_1} = \{v \in X_i \mid v \text{ not saturated by } M_{j_1}\}$ and $Y_{j_2} = \{v \in X_i \mid v \text{ not saturated by } M_{j_2}\}$. By these definitions we have that $M_{j_1}, M_{j_2}$ are a partition of $M$ and since $M$ is a matching in $G_i[V_i \setminus X] - U_i$ we conclude $X \subseteq Y_{j_1}, Y_{j_2}$ and $Y_{j_1} \cup Y_{j_2} = X$ and therefore $(Y_{j_1}, Y_{j_2}) \in \mathcal{S}_i(X)$. Thus, the matching $M$ is contained in the right hand side since it can be obtained by $M_{j_1} \cup M_{j_2}$. □

As a consequence of Lemma 3 we obtain the following theorem which shows how the tables corresponding to a join node can be constructed when the tables of its children are given.

**Theorem 8.** *Let $i \in I$ be a join node with children $j_1, j_2 \in I$. For $b \in \{0, \ldots, B\}$ the table $Q_i^b$ is correctly determined by the following algorithm.*

> **foreach** $a_i \in \mathcal{A}_i$ **do**
> |   $Q_i^b(a_i) = 0$;
> **end**
> **foreach** $b_1, b_2 \in \mathbb{N}$ *with* $b_1 + b_2 = b$ **do**
> |   **foreach** $a_{j_1} \in \mathcal{A}_{j_1}$, $a_{j_2} \in \mathcal{A}_{j_2}$ *with* $Q_i^{b_1}(a_{j_1}) = 1$ *and* $Q_i^{b_2}(a_{j_2}) = 1$ **do**
> |   |   An element $a_i \in \mathcal{A}_i$ is constructed as follows. For $X \subseteq X_i$ we set
> $$a_i(X) = \max\{a_{j_1}(X_{j_1}) + a_{j_2}(X_{j_2}) \mid (X_{j_1}, X_{j_2}) \in \mathcal{S}_i(X)\}$$
> |   |   $Q_i^b(a_i) = 1$;
> |   **end**
> **end**

### 4.5 Putting everything together

In a first step, the algorithm computes a *postorder tree walk* of $T$, which is an ordering of the nodes $I$, such that the position of a node is later in the ordering than any of its children. Such a walk can easily be computed in linear time. The algorithm goes through the nodes corresponding to the postorder tree walk and for every node $i \in I$, the corresponding tables $Q_i^b$, $b \in \{0, \ldots, B\}$ are determined with one of the four procedures explained above depending on the type of node $i$ (leaf node, introduce node, forget node or join node). As already noted at the beginning of Section 4 we have that $\nu_B(G) \leq K$ if and only if there is an entry $a_r \in \mathcal{A}_r^B$ with $a_r(\emptyset) \leq K$. This can easily be tested by going through all elements $a_r \in \mathcal{A}_r$ with $a_r(\emptyset) \leq K$ and checking whether $Q_r^B(a_r) = 1$.

We give a simple complexity analysis to show that the proposed algorithm runs in pseudo-polynomial time. For a leaf node $i \in I$, the tables $Q_i^b$ for $b \in \{0, \ldots, B\}$ are determined by the method described in Theorem 5. Since every leaf node $i$ satisfies $|X_i| = 1$, we have $|\mathcal{A}_i| \leq (C+1)^2$. Thus, computing all tables $Q_i^b$ for $b \in \{0, \ldots, B\}$ can be done in $O((B+1)(C+1)^2)$ time. For an introduce node $i \in I$ and a fixed $b \in \{0, \ldots, B\}$, the table $Q_i^b$ is determined by the algorithm described in Theorem 6. The initialization phase where the table is set to zero can be performed in $(C+1)^{(2^{k+1})}$ time and is not a bottleneck operation since it is dominated by the nested foreach loops that follow. Since $|E_i^v| \leq k$, the first foreach loop will be repeated at most $2^k$ times. Furthermore, as $|\mathcal{A}_j| \leq (C+1)^{(2^{k+1})}$, we have that the second foreach loop is called at most $(C+1)^{(2^{k+1})}$ times. Finally, for each $X \subset X_i$ (there are at most $2^{k+1}$ such subsets since $|X_i| \leq k+1$), the term $a_i(X)$ can be computed in $O(k)$ time because $|E_i^v \setminus U_i^v| \leq k$. Therefore, the total time needed for computing $Q_i^b$ for an introduce node $i$ is bounded by $O(2^k (C+1)^{(2^{k+1})} 2^{k+1} k)$. Since $k$ is a constant, this bound reduces to $O((C+1)^{2^{k+1}})$. Thus, for an introduce node $i$, the time needed for computing all tables $Q_i^b$ for $b \in \{0, \ldots, B\}$ is bounded by $O((B+1)(C+1)^{(2^{k+1})})$. For a forget node $i \in I$ and a fixed $b \in \{0, \ldots, B\}$, the table $Q_i^b$ is determined by the algorithm described in Theorem 7. The first as well as the second foreach loop is repeated at most $|\mathcal{A}_i| \leq (C+1)^{(2^{k+1})}$ times. Constructing $a^i$ in the second foreach loop can easily be done in $2^{k+1}$ time. For a forget node $i \in I$, the tables $Q_i^b$ for $b \in \{0, \ldots, B\}$ can therefore be constructed in $O((C+1)^{(2^{k+1})})$ time. Finally, for a join node $i \in I$ and a fixed $b \in \{0, \ldots, B\}$ we use the algorithm described in Theorem 8 for constructing the table $Q_i^b$. Again the first foreach loop is repeated at most $(C+1)^{(2^{k+1})}$ times and is not a bottleneck operation. The first of the nested foreach loops is repeated $b + 1 \leq B + 1$ times and the second one at most $|\mathcal{A}_{j_1}||\mathcal{A}_{j_2}| \leq (C+1)^{(2^{k+2})}$ times. For each $X \subset X_i$ we have to take the maximum over $|\mathcal{S}_i(X)|$ elements. Since $\mathcal{S}_i(X) \subseteq \mathcal{P}(X) \times \mathcal{P}(X)$ we have $|\mathcal{S}_i(X)| \leq (2^{k+1})^2$. For a join node $i \in I$, the time needed for computing all tables $Q_i^b$ for $b \in \{0, \ldots, B\}$ can therefore be bounded by $O(B^2(C+1)^{(2^{k+2})})$. Since the complexity bound we obtained for join nodes is the largest one of the four node types and since $|I| = O(n)$, we have that the total running time of the algorithm can be bounded by $O(nB^2(C+1)^{(2^{k+2})})$, which shows that the proposed algorithm runs in pseudo-polynomial time.

The presented algorithm can easily be modified to not only determine whether $\nu_B(G) \leq K$ but also to return an interdiction set $U$ with $\nu(G - U) \leq K$ if $\nu_B(G) \leq K$. As usual for dynamic programming this can be achieved by additional bookkeeping. For example by additionally storing for every table entry $Q_i^b(a_i)$ that is equal to one, an interdiction set that realizes $a_i$ in $G_i$ with respect to the budget $b$.

## 5 Conclusions

In this work, the matching interdiction problem was introduced as a natural problem arising when studying maximum matchings in the context of interdiction. Several complexity results were proven. In particular we showed that when the input is restricted to graphs consisting only of isolated edges, the matching interdiction problem is essentially a knapsack problem. Additionally, we proved that the matching interdiction problem is NP-complete on simple, bipartite graphs with unit edge weights and unit interdiction costs. This result implies strong NP-completeness of the matching interdiction problem. We presented a pseudo-polynomial algorithm for solving the matching interdiction problem on graphs with bounded treewidth. It is well known that many hard combinatorial problems like the maximum independent set problem can be solved efficiently on graphs with bounded treewidth. However, when dealing with interdiction problems we have to adapt the typical approach for building efficient algorithms on graphs with bounded treewidth because of the min-max nature of interdiction problems.

A possible continuation of this work would be to adapt the presented method to other interdiction problems on graphs with bounded treewidth. Additionally, it would be interesting to investigate in the complexity of matching interdiction problems on planar graphs. This direction of research is motivated by the network flow interdiction problem, which is a strongly NP complete problem in general but allows pseudo-polynomial algorithms for various subclasses of planar graphs [9, 12]. Furthermore, it would be interesting to investigate in approximation algorithms and in strong integer formulations of the matching interdiction problem.

# References

[1] H. L. Bodlaender. A tourist guide through treewidth. *Acta Cybernetica*, 11:1–21, 1993.

[2] H. L. Bodlaender. A linear-time algorithm for finding tree-decompositions of small treewidth. *SIAM Journal on Computing*, 25(6):1305–1317, December 1996.

[3] H. L. Bodlaender. A partial k-arboretum of graphs with bounded treewidth. *Theoretical Computer Science*, 209:1–45, 1998.

[4] H. L. Bodlaender and A. M. C. A. Koster. Combinatorial optimization on graphs of bounded treewidth. *The Computer Journal*, 2007.

[5] E. Boros, K. Elbassioni, and V. Gurvich. Transversal hypergraphs to perfect matchings in bipartite graphs: Characterization and generation algorithms. *Journal of Graph Theory*, 53(3):209–232, May 2006.

[6] M.-C. Costa, D. de Werra, C. Picouleau, B. Ries, and R. Zenklusen. Blockers and transversals. Technical report, 2008.

[7] M. R. Garey and D. S. Johnson. *Computers and Intractability; A Guide to the Theory of NP-Completeness*. W. H. Freeman & Co., New York, NY, USA, 1990.

[8] T. Kloks. *Treewidth: Computations and Approximations*, volume 842 of *Lecture Notes in Computer Science*. Springer Berlin / Heidelberg, 1994.

[9] C. A. Phillips. The network inhibition problem. In *STOC '93: Proceedings of the twenty-fifth annual ACM symposium on Theory of computing*, pages 776–785, New York, NY, USA, 1993. ACM Press.

[10] A. Schrijver. *Combinatorial Optimization, Polyhedra and Efficiency*. Springer, 2003.

[11] R. K. Wood. Deterministic network interdiction. *Mathematical and Computer Modeling*, 17(2):1–18, 1993.

[12] R. Zenklusen. Extensions to network flow interdiction on planar graphs. arxiv.org/abs/0801.1737, 2008.

# Appendix

*Proof of Lemma 1.* Let $X = \{x_1, \ldots, x_{p+q}\}$, $p + q > k$, and let $Y = \{y_1, \ldots, y_{\binom{k}{2}}\}$. Suppose for the sake of contradiction that the cardinality of a maximum matching $M$ in $G$ is $p \leq k$ and let us suppose w.l.o.g. that $M = \{\{x_1, y_1\}, \ldots, \{x_p, y_p\}\}$. To simplify notations we set $X_1 = \{x_1, \ldots, x_p\}$, $X_2 = \{x_{p+1}, \ldots, x_{p+q}\}$, $Y_1 = \{y_1, \ldots, y_p\}$ and $Y_2 = \{y_{p+1}, \ldots, y_{\binom{k}{2}}\}$. Note that since $k \geq 4$ and $p \leq k$, none of the sets $X_1, X_2, Y_1, Y_2$ is empty. As the vertices $X_2 \cup Y_2$ are not saturated by $M$, there is no edge $\{x, y\}$ in $E$ with $x \in X_2$ and $y \in Y_2$, otherwise $M$ would not be maximum. We define $I = \{i \in \{1, 2, \ldots, p\} \mid$ no alternating chain exists from any vertex in $X_2$ to $x_i\}$, where an alternating chain is a chain whose edges are alternately in $E \setminus M$ and $M$.

Every vertex $y_i \in Y$ with $i \in I$ has both neighbors in $\{x_i \in X \mid i \in I\}$, otherwise we could find an alternating chain from a vertex of $X_2$ to $x_i$. Furthermore at most $\binom{|I|}{2}$ vertices in $Y$ can have both of their neighbors in $\{x_i \in X \mid i \in I\}$ as otherwise there would be two vertices in $Y$ having the same neighbors, thus violating the condition that $G$ contains no $C_4$. Notice that all vertices in $Y_2$ have both neighbors in $\{x_i \in X \mid i \in I\}$ since otherwise we would have an augmenting chain $W$, i.e., an alternating chain where both of its endpoints are not saturated by $M$. However this contradicts maximality of $M$ since $M \setminus W \cup W \setminus M$ would be a matching in $G$ with larger cardinality than $M$. Hence, all vertices in $Y_2 \cup \{y_i \in Y \mid i \in I\}$ have both neighbors in $\{x_i \in X \mid i \in I\}$ and since there are not two vertices in $Y_2 \cup \{y_i \in Y \mid i \in I\}$ with the same neighbors we have

$$|Y_2 \cup \{y_i \in Y \mid i \in I\}| = \binom{k}{2} - p + |I| \leq \binom{|I|}{2} . \tag{1}$$

Reformulating and developing the above inequality we obtain the following contradiction.

$$\binom{k}{2} \leq \binom{|I|}{2} - (|I| - 1) + p - 1 = \binom{|I| - 1}{2} + p - 1$$
$$< \binom{p-1}{2} + p - 1 = \binom{p}{2} \leq \binom{k}{2}$$

The first inequality is a simple reformulation of inequality (1). The first equality is due to the relation $\binom{i}{2} - (i - 1) = \binom{i-1}{2}$, $\forall i \in \{2, 3, \ldots\}$ and the fact that $I$ contains at least two elements because the vertex $y_{p+1}$ has its two neighbors in $X_1$ which must be in the set $\{x_i \in X \mid i \in I\}$, as otherwise we have an augmenting chain. The strict inequality is implied by the fact that $|I| < p$ because the vertex $x_{p+1}$ has at least one neighbor $y_i$ in $Y_1$, implying that $i \notin I$ as the chain along the vertices $(x_{p+1}, y_i, x_i)$ is alternating. The last equality and inequality follow from the binomial relation already highlighted and the relation $p \leq k$. □